# Quantum Fluctuations drive Emerging Bosonic Quantum Phases in a Moiré Flatband Superconductor

Yu Hong[1,2#], Haiming Sun[3#], Qixing Zhang[1,2], Le Liu[1,2], Zitian Pan[1,2], Jundong Zhu[1,2], Jingwei Dong[1,2], Kenji Watanabe[4], Takashi Taniguchi[5], Luojun Du[1,2], Dongxia Shi[1,2], Adrian Bachtold[6], Haiwen Liu[3*], Guangyu Zhang[1,2*], Wei Yang[1,2*]

[1] *Beijing National Laboratory for Condensed Matter Physics and Institute of Physics, Chinese Academy of Sciences, Beijing 100190, China*

[2] *School of Physical Sciences, University of Chinese Academy of Sciences, Beijing 100190, China*

[3] *Center for Advanced Quantum Studies, Department of Physics, Beijing Normal University, Beijing 100875, China*

[4] *Research Center for Electronic and Optical Materials, National Institute for Materials Science, 1-1 Namiki, Tsukuba 305-0044, Japan*

[5] *Research Center for Materials Nanoarchitectonics, National Institute for Materials Science, 1-1 Namiki, Tsukuba 305-0044, Japan*

[6] *ICFO - Institut de Ciencies Fotoniques, The Barcelona Institute of Science and Technology, Castelldefels 08860, Barcelona, Spain*

*# These authors contribute equally.*

**Corresponding Authors (emails): haiwen.liu@bnu.edu.cn; gyzhang@iphy.ac.cn; wei.yang@iphy.ac.cn*

## ABSTRACT

Quantum fluctuations play a central role in superconductors by mediating the interactions that promote electron pairing and controlling the phase coherence required for macroscopic condensation. Thus far, understanding the interplay between quantum fluctuations and superconductivity remains challenging in unconventional superconductors. Here, we establish quantum phase fluctuations as the organizing principle linking superconductivity and Bose-insulating quantum criticality in a moiré flatband superconductor of alternating twisted trilayer graphene (ATTG). Superconductivity is optimal near half-filling of the moiré flatband and gives way to a density-wave-like resistance peak at high temperatures, in which we observe distinct temperature scales for the pairing (~6 K) and the condensation (~1 K). We observe compelling evidence of a bosonic superconducting quantum phase transition, driven by both out-of-plane and in-plane magnetic fields, with a universal critical resistivity of $R_c \approx 0.62R_Q$ and finite-temperature scaling of $z\nu_B \approx 2.3$, where $R_Q = h/4e^2$ is the quantum resistance for bosons. Strikingly, the observed bosonic insulator shows a nearly quantized resistivity of $R_Q$ that saturates at low temperatures and high magnetic fields, together with a zero-bias peak, a nearly zero Hall resistance, and a logarithmic dependence at higher temperatures. Both the superconductor and the bosonic insulator could be driven into the same critical quantum fluctuation regime dynamically under a large bias. Our results establish moiré flatband as a clean platform for studying unconventional superconductivity and bosonic quantum criticality governed by quantum fluctuations.

## I. INTRODUCTION

Quantum fluctuations and electron correlation are essential for understanding unconventional superconductors intertwined with competing or cooperative orders [1-6]. Moiré superlattices have been demonstrated as an ideal platform for investigating novel superconductivity and strong correlation in the flatband limit [7-22]. The observations of a domed superconducting structure near correlated insulators [10-13], strongly coupled superconductivity with a large ratio of $k_B T_c$ over Fermi energy [14,15,17], and spectroscopic signatures of nodal superconductivity [14,23] all bear close resemblance to the high-$T_c$ superconductors [3,4,6]. In addition, the nontrivial topology endorses a new ingredient [24], and the quantum metric of moiré flatbands is believed to play a pivotal role in determining superfluid stiffness [25-29]. Despite the progress, however, it remains a major challenge to fully understand the nature of superconductivity in moiré flatbands [13,14,30-32]. The key lies in the complex order parameters $\Delta e^{i\Phi}$ of the superconductor, where superconductivity emerges when Cooper pairs (bosons due to bound electron pairs) undergo a global phase condensation. For superconductors in two dimensions, phase fluctuations and related quantum phase transitions are of vital importance [1,2,33]. Subject to sufficiently strong fluctuations or disorder, bosonic superconductor-insulator quantum phase transitions can occur with a critical sheet resistance close to the quantum resistance $R_Q = h/4e^2$, where $h$ and $e$ are Planck's constant and the electron charge.

Superconductor–insulator transition (SIT) driven by magnetic fields is a prototypical two-dimensional quantum phase transition and a long-standing problem of how superconductivity is destroyed at low temperature [34-43]. Two limiting scenarios are usually distinguished. In a conventional pair-breaking, or fermionic, scenario, increasing the field suppresses the pairing amplitude $|\Delta|$ and drives the system toward a fermionic normal or insulating state. In a bosonic scenario, by contrast, local pairing survives while global phase coherence of $e^{i\Phi}$ is lost, producing a Bose-insulating regime of localized paired carriers [34,40]. The two scenarios differ in what the magnetic field destroys: the pairing amplitude or the global phase. The critical resistance is often of the order of $R_Q = h/4e^2$, but need not be exactly universal, since normal channels, dissipation, and inhomogeneity may contribute in parallel. In conventional thin films and oxide interfaces [35,36], this distinction is often obscured by uncontrolled chemical disorder, emergent granularity, and fermionic dissipation, motivating the search for a clean, tunable moiré platform.

In this work, we provide new insights into the unconventional superconductivity and bosonic superconducting quantum phase transition in a moiré flatband superconductor. We focus on alternating twisted trilayer graphene (ATTG) with $\theta \approx 1.3°$, in which we observe clean and robust 2D superconductivity with the Berezinskii-Kosterlitz-Thouless (BKT) transition. Being a flatband system in the intermediate coupling regime with $U/t$ ~1, where $U$ and $t$ denote the Coulomb interaction and kinetic energy, respectively, ATTG with a small twist angle could host a rich phase diagram with competing orders and instabilities. Here, superconductivity is observed optimally doped with the highest $T_c$ at half filling of the flatband, while it develops into a density wave state featured as a resistance peak at high temperatures, suggesting an intimate correlation between superconductivity and density wave. Moreover, we observe compelling evidence of bosonic superconductor-insulator quantum phase transition by applying magnetic fields, regardless of the field direction. The observed critical resistance of 0.62 $R_Q$ is close to the quantum resistance ($R_Q$) expected for a bosonic transition [34], and finite-temperature scaling (FTS) gives a critical exponent $z\nu_B = 2.3$, in line with the 2D quantum percolation model. The insulating state exhibits logarithmic temperature dependence $R \sim \ln T$, together with an unconventional zero Hall resistance, suggestive of tightly bound bosons without global phase coherence. Similar to the superconducting state, the field-induced insulating state also exhibits pronounced nonlinear transport at low temperatures, while both sides recover a common linear dynamic transport regime above $\sim 6$ K or under sufficiently large dc bias. This continuity indicates that the superconducting and insulating branches are organized from the same paired degrees of freedom, but with different phase organization. We describe the transition as a magnetic field-driven collapse of a phase-coherent Josephson network where the competition of Josephson coupling and Coulomb interaction is essential in the Bose-Hubbard model [44-46]. Our observations highlight the crucial role of phase fluctuations and Andreev coherence in organizing superconductivity and Bose-insulating quantum criticality in a clean moiré flatband system.

## II. RESULTS

### A. Unconventional superconductivity and quantum fluctuations

ATTG is encapsulated between two hBN flakes and patterned into a dual-gate geometry, which permits independent control of the carrier density $n$ and the electric displacement field $D$ (see Supplementary Material Note 1 [47] for sample information). ATTG features a coexistence of moiré flatbands and a dispersive Dirac cone [48-51], and the displacement field hybridizes the two bands and leads to enhanced superconductivity at finite $D$. Figure 1(a) displays plots of sheet resistance $\rho_{xx}$ versus $\nu$ for different temperatures at $D$ = -0.6 V/nm, which is taken along the red dashed line in Fig. 1(c), corresponding to an optimal displacement field for superconductivity. It reveals a robust superconducting state with zero resistance between $\nu = 2$ and $\nu = 3$ at $T = 60$ mK and its transition to a resistive state with a peak at $\nu = 2.5$ at high temperatures $T > 2$ K. Figure 1(b) shows representative $\rho_{xx}$ (T) curves at different fillings. In the underdoped regime ($\nu < 2$), it shows a direct evolution from a metallic state with Fermi liquid behavior, $\rho_{xx} \propto T^2$ at $T > T^*$, to the onset of superconductivity when cooled down to $T = T^*$ ($T^* \sim 0.35$ K for $\nu = 1.62$). In the optimally doped regime ($\nu \sim 2.5$), it shows metallic behavior at $T > T^* \sim 6$ K and then a plateau or weak insulating behavior before entering the superconducting regime at $T < T_c \sim 1.2$ K. The observed $T_c$ agrees with the Berezinskii–Kosterlitz–Thouless (BKT) transition temperature [52,53] $T_{BKT} = 1.05$ K in Fig. 1(e), indicating a development of long-range phase coherence of Cooper pairs in two dimensions below $T_c$.

We can further map out the superconducting phase diagram by including $T_c$ and $T^*$ in Fig. 1(d). The superconductivity is marked by the zero resistance that spans between $\nu = 2$ and 3, revealing a domed structure defined by $T_c$ (pink dots), and the metallic behavior is resumed at high temperatures defined by $T^*$ (purple dots). In the underdoped region ($\nu < 2$), doping induces a Fermi liquid-to-superconductor transition. In the overdoped regime ($\nu > 3$), superconductivity is suppressed upon doping, leading to a superconductor-to-insulator transition.

The coexistence of a high-temperature resistance peak and a low-temperature superconductor with maximized $T_c$ at $\nu = 2.5$ is striking. The resistivity peak near $\nu = 2.5$ can be attributed to symmetry breaking at half filling of the moiré conduction flatband [30], or equivalently to a metallic density wave commensurate with the moiré superlattice [54,55] that signals electronic reconstruction or competing correlated order [5]. A competing density wave instability can reduce phase stiffness and provide a natural background for separating local pair formation from global phase ordering[2]. Similar resistance peaks also appear at $D = 0$ and in a separate device (Supplemental Material Note 2 [47]). The observations suggest a picture of unconventional superconductivity where fluctuations play a crucial role [56,57]. Also note that a Mott insulator to gossamer superconductivity transition could be expected at half filling for a moderate Coulomb interaction in the Hubbard model [58].

The observation of two characteristic temperatures, $T^*$ and $T_c$, suggests a scenario of pairing and condensation at high and low temperatures, respectively, reminiscent of the pseudogap phase in unconventional high-$T_c$ superconductors [59-61]. This separation between $T^*$ and $T_c$ establishes the central phase-fluctuation premise of our study: the fragile low-energy degree of freedom is the global phase coherence of preformed pairs, rather than the local pairing amplitude itself. The weak insulating behavior or resistance plateau between two temperatures points to a fluctuating correlated regime, in which competing electronic order and low phase stiffness coexist before the establishment of global superconducting coherence. Such a scenario is further supported by the observation of a saturated sheet resistance of 6.8 kΩ at $\nu = 2.5$ when superconductivity breaks down under finite magnetic fields [Fig.1(f)]. Strikingly, the saturated $\rho_{xx}$ is close to $R_Q = h/4e^2$, consistent with transport dominated by quantum fluctuations of paired bosons.

### B. Bosonic superconducting quantum phase transition

To elucidate the unconventional nature and the order parameters of superconductivity, we study superconducting quantum phase transitions at the optimal doping by applying magnetic fields. Fig. 2(a) shows the temperature dependence of the $\rho_{xx}$ for various $B_\perp$ at $\nu = 2.5$ and $D = -0.6$ V/nm. At low temperatures $T < 2$ K, the superconductivity is rapidly suppressed when introducing a small $B_\perp$, and it turns into an insulating state ($d\rho_{xx}/dT < 0$) under high magnetic fields $B_\perp > 0.4$ T. To better reveal the critical behavior of the magnetic field-driven SIT, we replot the data as $\rho_{xx}$ versus the magnetic field at different temperatures in Fig. 2(b). Interestingly, $\rho_{xx}(B)$ curves intersect at one single point, corresponding to a critical magnetic field $B_c$ = 0.345 T and a critical resistance $R_c$ = 3980 Ω that is insensitive to temperature changes.

The observed magnetic field-induced SIT is unconventional in several ways. To begin with, the critical sheet resistance is about $0.62R_Q$, of the order of the quantum resistance expected for a bosonic (dirty-boson) SIT [34,62]; its precise value is not expected to be strictly universal in the presence of parallel conduction channels and dissipation [40]. Note that SITs were reported in graphene-based moiré superlattices [63,64]; however, the transitions are either obscured by the strong Mott-like gap at half filling or strongly influenced by the fermionic conduction, leaving the bosonic SIT unexplored. Aside from the critical resistance, we observe more experimental facts to support the bosonic SIT picture. For instance, the Hall resistance is around zero during the magnetic field-induced phase transition [inset of Fig. 2(b)], which further supports the scenario of localized bosons. In addition, the insulator shows strong nonlinearity, and the resistivity at zero bias follows a logarithmic decay with temperature, i.e., $\rho_{xx} \propto -\ln T$ [inset of Fig. 2(a)], unlike the thermally activated behavior [65]. Although weak localization (WL) of electrons in 2D systems also gives rise to a logarithmic temperature dependence of resistance, it is readily suppressed by magnetic fields [66]; the persistence of logarithmic dependence at $B_\perp$ = 1 T therefore rules out WL as the origin of the insulating state. In contrast, phase fluctuations can logarithmically renormalize the inter-puddle tunneling conductance [67], which appears as an approximately logarithmic increase of resistance over a finite temperature window. Moreover, the logarithmic dependence can persist up to $T = T_\Delta \sim 6$ K, above which metallic behavior is observed, which matches well with $T^* \sim 6$ K under zero magnetic fields, indicative of the same origin. The distinct critical temperature between the superconductor and the insulator agrees with superconductivity dominated by phase fluctuations. To sum up, these observations systematically establish a scenario of a bosonic superconductor-insulator quantum phase transition driven by magnetic fields.

To further confirm the quantum critical nature of the bosonic SIT, we perform finite temperature scaling (FTS) analysis of $\rho_{xx}(B)$ curves in the vicinity of quantum critical point (QCP) within a temperature window of 0.3 to 1.1 K [Fig. 2(c)]. Based on FTS, the resistivity follows a universal scaling law near QCP, i.e., $\rho_{xx}(B,T) = R_c \mathcal{F}(|B - B_c|T^{-1/z\nu_B})$, where $\mathcal{F}(x)$ is the universal scaling function with $\mathcal{F}(0) = 1$, $\nu_B$ and $z$ are the critical exponents of the diverging spatial and dynamic correlation length [34,62,68]. The critical exponent product of $z\nu_B$ is referred to as the universality class of QPT, which ignores microscopic details and depends only on the system's effective dimensionality, disorder, and the range of interactions. The result of FTS is shown in Fig. 2(c). All the data are collapsed into two distinct branches, one for $B > B_c$ and the other $B < B_c$, thereby capturing the key evidence of QPT. Physically, the magnetic field progressively suppresses the phase-coherent Andreev links between local pairing regions. The critical point is reached when these coherent links no longer form a sample-spanning network. Thus, $z\nu_B \simeq 2.3$ reflects a quantum-percolative collapse of the phase-coherent Cooper-pair transport network, where strong phase quantum fluctuations lead to localized bosons without global phase coherence [68,69].

We reproduce the SIT scaling behavior with the same critical exponents across different contact pairs in the same sample, as well as in a separate sample with a twist angle of 1.26°, demonstrating the universality of the SIT transition in small-angle ATTG (Supplemental Material Note 3 [47]). Note that we also observe a doping-induced SIT, as shown in Supplemental Material Note 4 [47]. FTS of doping-driven SIT yields a critical exponent product $z\nu_B = 1.5$ and a critical resistivity of $R_c \approx 0.62\ R_Q$. The exponent product, which differs from 2.3 in the magnetic field-driven SIT, points to a distinct microscopic mechanism for doping-driven SIT, whereas the identical critical resistivity may signal the universality of the fluctuation-dominated QPT in this system.

### C. Nonlinear critical behaviors in QPT

We further provide new insight into the bosonic superconducting quantum phase transition by studying the nonlinearity in depth. Fig. 3(a) shows plots of differential resistivity $\mathrm{d}V/\mathrm{d}I$ against current at different magnetic fields. At $B = 0$, it features a zero $\mathrm{d}V/\mathrm{d}I$ at small current and a sharp $\mathrm{d}V/\mathrm{d}I$ peak at a critical current $I_c \sim 85$ nA, and then it ends up with a nearly constant $\mathrm{d}V/\mathrm{d}I$ at large current densities $I > I_c$, marking the breakdown of superconductivity and the global phase coherence where the vortex-antivortex pairs are expected to proliferate [70]. With the increase of $B$ approaching $B_c$, the amplitude of $\mathrm{d}V/\mathrm{d}I$ peak and $I_c$ are rapidly suppressed, while the $\mathrm{d}V/\mathrm{d}I$ at $I > I_c$ remains unchanged. The emergence of a narrow zero-bias resistivity peak for $B > B_c$ indicates a strong suppression of low-energy tunneling. Rather than a hard single-particle gap, this feature is naturally interpreted as a zero-bias anomaly (ZBA) produced by phase fluctuations in a phase-disordered localized-pair state [67]. The temperature-dependent nonlinear response reveals the simultaneous weakening of the superconducting nonlinear coherence feature and the insulating zero-bias anomaly at $B > B_c$. Notably, the zero-bias anomaly is strongest near the displacement field with the highest $T_c$ and $I_c$ and the shortest coherence length $\xi$ (Supplemental Material Note 5 [47]). These observations point to an intrinsic connection between superconductivity and quantum phase fluctuations in ATTG.

Interestingly, under high magnetic fields and low temperatures, the zero-bias resistivity of the insulator is saturated at around 6.8 kΩ, close to the quantum resistance for bosons $R_Q = h/4e^2$. As temperature increases, $I_c$ and the corresponding $\mathrm{d}V/\mathrm{d}I$ peaks decrease in the superconducting state [Fig. 3(d)], while the zero-bias $\mathrm{d}V/\mathrm{d}I$ peaks become lower and broader in the insulating state [Fig. 3(c)]. The low-temperature saturation of the zero-bias resistivity, as shown in Supplemental Material Note 6 [47], indicates that the insulating branch remains close to a quantum-critical localized-pair regime with finite inter-puddle tunneling. This behavior is consistent with a weak Bose-insulating regime rather than an idealized fully localized dirty-boson limit [67]. The observation of soft low-energy tunneling suppression resembles that of granular superconductor [36], where the QPT involves localization of bosonic Cooper-pair islands.

Note that we reproduce the same bosonic superconducting quantum phase transition under in-plane magnetic fields. Fig. 3(b) shows typical nonlinear behaviors in the insulating state at $B_{\parallel} > 4.6$ T, similar to those in Fig. 3(a) under out-of-plane magnetic fields. In addition, FTS analysis reveals the identical quantum phase transition for both in-plane and out-of-plane magnetic fields, evidenced by the same critical resistance $R_c$ and the same critical exponent $z\nu_B = 2.3$ (Supplemental Material Note 7 [47]), which further demonstrates a universal quantum criticality driven by phase quantum fluctuations. The microscopic coupling of $B_{\parallel}$ and $B_{\perp}$ to the device can be different: $B_{\perp}$ directly introduces orbital frustration and vortices, whereas $B_{\parallel}$ primarily couples through Zeeman energy. The identical scaling and nonlinear evolution indicate that both field directions flow to the same low-energy phase-fluctuation regime.

Last but not least, we observe a universal dynamic transport behavior when the system deviates from nonlinearity and restores linearity at large currents. As shown by triangles in Fig. 3(e), for both the superconductor and the insulator subjected to sufficiently large DC currents, $\mathrm{d}V/\mathrm{d}I$ is saturated at $0.62R_Q$, and it remains unchanged until reaching a critical temperature of ~6 K, above which it recovers a metallic behavior. The convergence of the superconducting and insulating branches at large bias indicates that the two sides share the same paired microscopic degrees of freedom. The magnetic field changes the macroscopic phase organization of these pairs, from a phase-coherent network to a phase-disordered localized-pair network, rather than converting them into ordinary fermionic quasiparticles.

Taken together, these observations provide compelling evidence for a scenario in which optimally doped superconductivity at half filling is dominated by phase quantum fluctuations. Bosonic electron pairing sets in at a critical temperature of ~6 K, at which the transport is in the quantum critical fluctuation regime (with a critical resistivity close to $R_Q$) where Cooper pairing competes against the density wave, and boson condensation with global phase coherence sets in at a much lower BKT temperature of ~1 K. These results motivate a phase-network description of the field-driven SIT. The superconducting branch corresponds to a percolating network of phase-coherent paired regions, while the insulating branch corresponds to localized paired carriers without global phase coherence. The magnetic field suppresses the low-energy coherent Josephson network and largely influences the phase of preformed pairs. Upon increasing temperature or applying a large current above the critical point, the condensed phase becomes unstable towards bosonic excitations with a universal critical resistivity set by quantum fluctuations.

### D. Microscopic process of bosonic SIT

We now turn to a detailed analysis of the nonlinear $\mathrm{d}V/\mathrm{d}I$ response induced by current $I_{dc}$ to better understand the QPT process. We first establish a phenomenological model for quantum fluctuations-driven SIT in a moiré system [Fig. 4(a)]. At low temperature and zero magnetic field, local paired carriers develop global phase coherence and form a dissipationless BKT superconductor. When a magnetic field is applied, the local pairing can remain robust because of the strong correlations in the moiré flatband, but the global phase organization is progressively disrupted. For perpendicular fields, vortices and orbital phase frustration weaken the coherent links; for in-plane fields, the Zeeman effect provides a different microscopic route to the same low-energy outcome. Thus, the system should not be viewed as a structurally granular film, but as an emergent moiré network of local paired puddles embedded in a conducting background. In the superconducting branch, these puddles are connected by phase-coherent Josephson links; in the insulating branch, the same paired carriers survive but lose global phase coherence and become localized. Under sufficiently large dc bias, or above the common temperature scale of about 6 K, the low-energy nonlinear structures on both sides are washed out, and the system enters a dynamic paired transport regime.

Microscopically, the experimental observations [Fig. 4(b)] can be captured by taking into account the Andreev dissipation in the Bose-Hubbard model. Because the local paired puddles are coupled to a conducting moiré background, their phase dynamics are protected by Andreev subgap dissipation, denoted by $\mathcal{G}_A$. In the Larkin-Feigel'man picture [71], this dissipative protection renormalizes the bare charging energy into an effective value $E_C^* = E_C \exp\left(-\frac{\pi\mathcal{G}_A}{4}\right)$. Magnetic field therefore drives the SIT through two cooperative effects: it weakens the coherent inter-puddle coupling $J_{ij}^{\mathrm{coh}}$ and, at the same time, suppresses the Andreev dissipative protection $\mathcal{G}_A$, thereby restoring $E_C^*$ and enhancing quantum phase fluctuations. Based on this picture, the low-energy transport is governed by two processes: (i) Andreev coherent tunneling via superconducting fluctuations [71,72], contributing a conductance $G_{\mathrm{coh}}$ and (ii) Coulomb-blockade dominated insulating paths [67], described by $G_{\mathrm{inc}}$. The total conductance is then $G_{total} = G_{\mathrm{coh}} + G_{\mathrm{inc}}$ (see Supplemental Material [47] for the detailed fitting procedure and a discussion of the Larkin–Feigel'man Andreev proximity channel [71,72] and the Coulomb-blockade channel [67]). The simulation results are shown in Fig. 4(c), exhibiting a qualitative agreement with the experimental data. In the high-field insulating regime, the spectra are well fitted by the insulating channel $G_{\mathrm{inc}}$ alone, and the increase of the zero-bias anomaly strength coefficient $C^*$ (defined in Supplemental Material [47]) with magnetic field shows that phase fluctuations and the ZBA become stronger as the system moves deeper into the localized-pair regime. Near and below $B_c$, the low-energy contribution $G_{\mathrm{coh}}$ from coherent channel cannot be neglected [Fig. 4(c)]. The critical field is reached when the phase-coherent Andreev links no longer form a sample-spanning network; the exponent product $z\nu_B \simeq 2.3$ is the scaling fingerprint of this quantum-percolative loss of global phase coherence, rather than a homogeneous pair-breaking transition. Once the coherent network loses global connectivity, the insulating branch is governed by Coulomb-renormalized incoherent tunneling. The low-energy zero-bias peak should therefore be understood not as a hard single-particle gap, but as a zero-bias anomaly: a tunneling event changes the local charge of a moiré puddle, the surrounding screening cloud must readjust, and the resulting phase fluctuations suppress the low-energy tunneling density of states. The nonlinear response, together with the two-process analysis, provides powerful evidence for phase fluctuation-driven SIT in ATTG system.

## III. CONCLUSIONS

In conclusion, our work identifies the importance of quantum fluctuations in determining the unconventional superconductivity and quantum phase transitions in two-dimensional moiré flatbands. We extract $R_c = 0.62R_Q$ and $z\nu_B = 2.3$ from finite-temperature scaling, supporting a quantum-percolative collapse of phase-coherent Cooper-pair transport rather than homogeneous pair breaking. The near-linear $R$–$\ln T$ dependence, zero-bias anomaly, suppressed Hall response, and large-bias convergence of the superconducting and insulating branches are consistently explained by a localized-pair Bose-insulating transport regime. Microscopically, increasing magnetic field collapses a coherent Josephson network by weakening inter-puddle coherence and Andreev dissipation, thereby restoring the effective charging energy and enhancing quantum phase fluctuations. Above the critical field, local pairs survive but lose global coherence, and their transport is governed by Coulomb-renormalized tunneling and low-energy suppression. These results establish ATTG as a pristine platform for realizing bosonic quantum criticality, bridging flatband superconductivity and quantum phase fluctuations in two dimensions.

**Acknowledgments**

We thank Xi Dai, KT Law (HKUST), Yi Zhou, Kun Jiang, Yifeng Yang, and Shiliang Li (IOP-CAS) for fruitful discussions. We acknowledge support from the Quantum Science and Technology-National Science and Technology Major Project (Grant No. 2025ZD0300500), Beijing Natural Science Foundation (JQ25003), the National Key Research and Development Program (Grant No. 2024YFA1410400), and the Natural Science Foundation of China (NSFC, Grant No. 61888102). K.W. and T.T. acknowledge support from the Elemental Strategy Initiative conducted by the MEXT, Japan (Grant No. JPMXP0112101001), JSPS KAKENHI (Grant Nos. 19H05790, 20H00354 and 21H05233) and A3 Foresight by JSPS.

**Data Availability Statement**

The data that support the findings of this study are openly available.

**Author contributions**

W.Y. and G.Z. supervised the project. W.Y. and Y.H. designed the experiments. Y.H. and Q.Z. fabricated the devices with the help of L.L., Z.P., J.Z., J.D.; Y.H. performed the magneto-transport measurement. H.S. and H.L. developed the theoretical model and the simulations. K.W. and T.T. provided hexagonal boron nitride crystals. Y.H., H.S., X.D., A.B., H.L., G.Z., and W.Y. analyzed the data. Y.H., H.S., G.Z., H.L., and W.Y. wrote the paper with the input from all the authors.

**Competing interests**

The authors declare no competing interests.

# Figures and Figure Captions

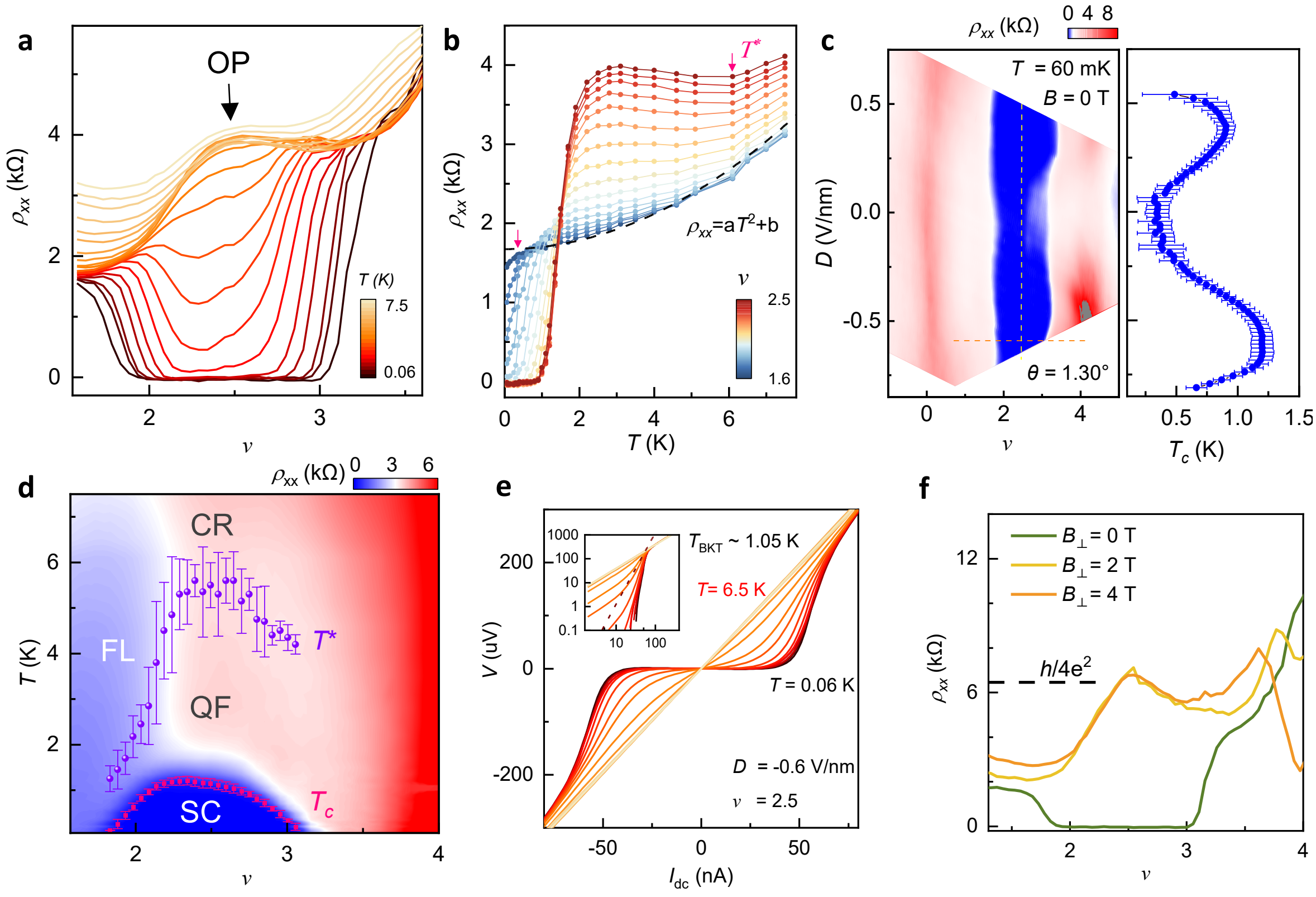

**Fig. 1 Unconventional superconductivity in alternating twisted trilayer graphene. (a)** $\rho_{\mathrm{xx}}$ as a function of moiré filling factor $v$ from 7.5 K to 60 mK at $D$ = -0.6 V/nm. **(b)** $\rho_{\mathrm{xx}}$ as a function of $T$ from $v$ = 1.62 to 2.5, at fixed $D$ = -0.6 V/nm. The dashed line denotes Fermi liquid behavior, with a and b as two fitting parameters. The purple arrows indicate the onset temperature $T^*$ of the deviation from metallic behavior. **(c)** Left, $\rho_{\mathrm{xx}}$ as a function of $v$ and $D$ at $T$ = 60 mK. The white and orange dashed lines denote $v$ = 2.5 and $D$ = -0.6 V/nm, respectively. Right, superconducting critical temperature $T_{\mathrm{c}}$ as a function of $D$ at $v$ = 2.5. $T_{\mathrm{c}}$ is extracted at 20% of the normal-state resistance (see Supplemental Material Note 1 [47]), with the error bar determined from the 10%-30% range. **(d)** $v$-$T$ phase diagram at $D$ = -0.6 V/nm. The pink and purple dots denote $T_{\mathrm{c}}$ and $T^*$, respectively. FL, Fermi liquid; CR, critical region; QF, quantum fluctuation region; SC, superconducting region. **(e)** $V$ versus $I$ from 60 mK to 6.5 K. Inset, zoom-in figure in log-log coordinates. The brown dashed line indicates the $V \sim I^3$ for the BKT transition. **(f)** $\rho_{\mathrm{xx}}(v)$ for different out-of-plane magnetic field $B_{\perp}$ = 0, 2 and 4 T at $T$ = 60 mK.

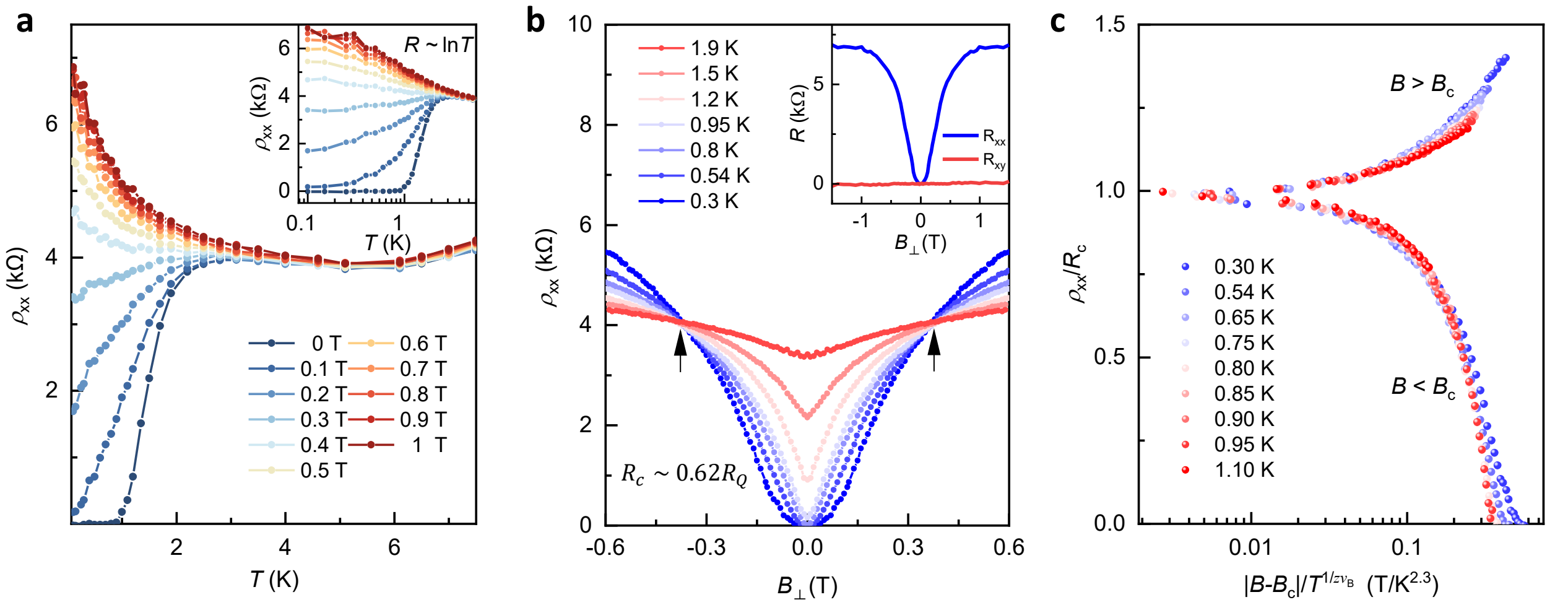

**Fig. 2 Bosonic superconducting quantum phase transition. (a)** $\rho_{xx}$ as a function of $T$ upon increasing out-of-plane magnetic field ($B_\perp$) from 0 T to 1 T. Inset, zoom-in figure in semi-log coordinate. **(b)** $\rho_{xx}$ as a function of $B_\perp$ from $T = 0.3$ K to 1.9 K. Inset, $R_{xx}$ and $R_{xy}$ versus $B_\perp$ at $T = 60$ mK. Black arrows show the quantum critical points. **(c)** The finite temperature scaling (FTS) analysis showing $\rho_{xx}/R_c$ as a function of scaling variable $|B - B_c|/T^{1/z\nu_B}$ over the temperature range 0.3 K-1.1 K, where $R_c = 0.62R_Q$ is the critical sheet resistance, $B_c$ is the critical field, $z\nu_B = 2.3$ is the scaling exponent.

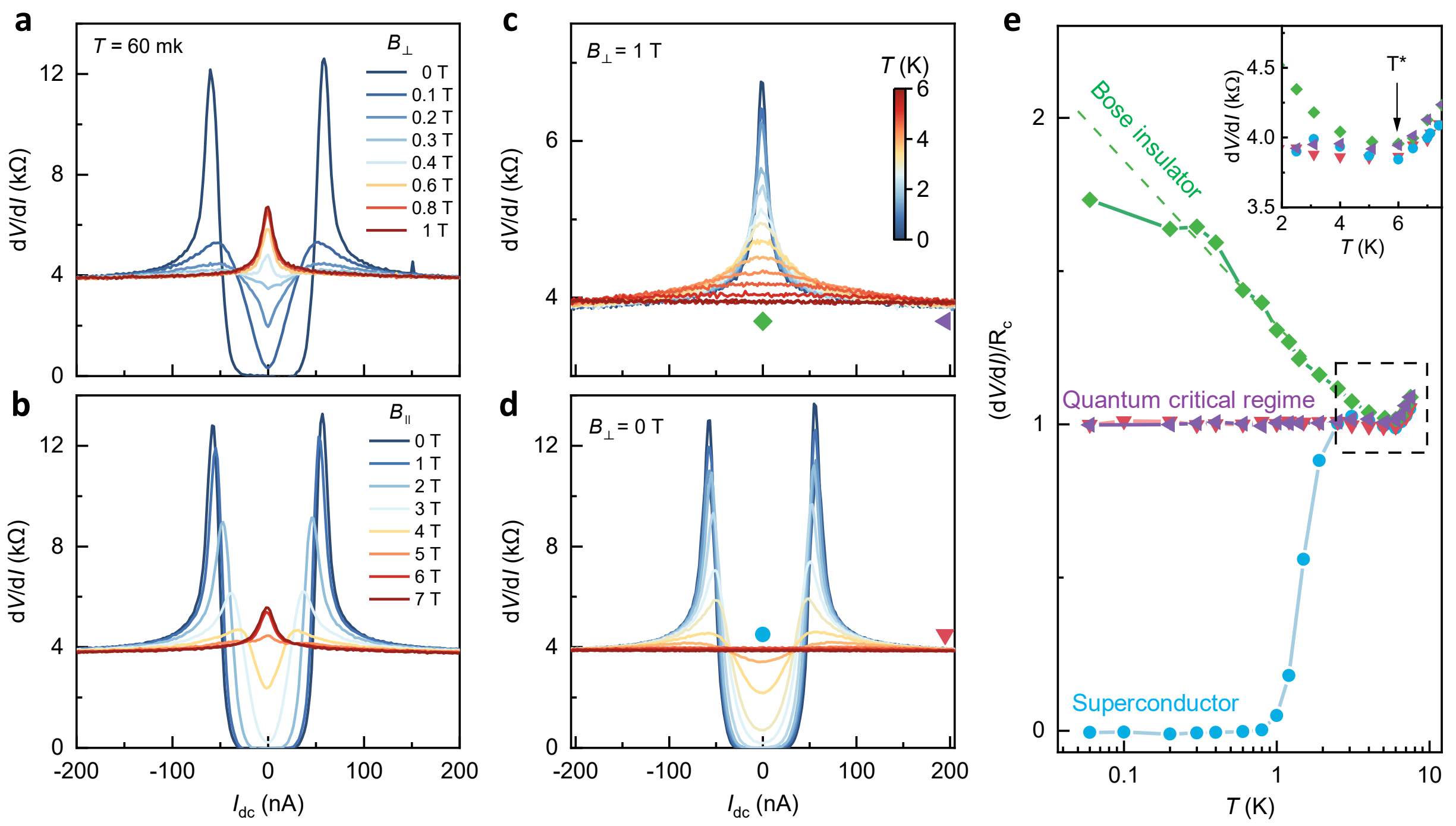

**Fig. 3 Nonlinear transport behavior of superconductor and Bose insulator. (a)** Differential sheet resistance d$V$/d$I$ as a function of the dc bias current $I_{dc}$ for out-of-plane magnetic field $B_{\perp}$ from 0 T to 1 T at $T$ = 60 mK. **(b)** d$V$/d$I$ as a function of $I_{dc}$ for in plane $B_{\parallel}$ from 0 T to 7 T at $T$ = 60 mK. **(c)(d)** d$V$/d$I$ as a function of $I_{dc}$ from $T$ = 60 mK to 6 K at $B_{\perp}$ = 1 T (the insulating side) and $B_{\perp}$ = 0 T (the superconducting side), respectively. **(e)** Normalized differential resistance, (d$V$/d$I$)/$R_c$, as a function of $T$ on a logarithmic scale, extracted from **(c)** and **(d)** at $I_{dc}$ = 0 nA (superconductor, blue; insulator, green) and $I_{dc}$ = 200 nA (dynamic superconductor, red; dynamic insulator, purple). $R_c$ is the critical resistance at the quantum critical point. The Inset in **(e)** shows d$V$/d$I$ at $T$ > 2 K on a linear scale. The arrow marks $T^*$ ≈ 6 K, where the states regain metallic behavior.

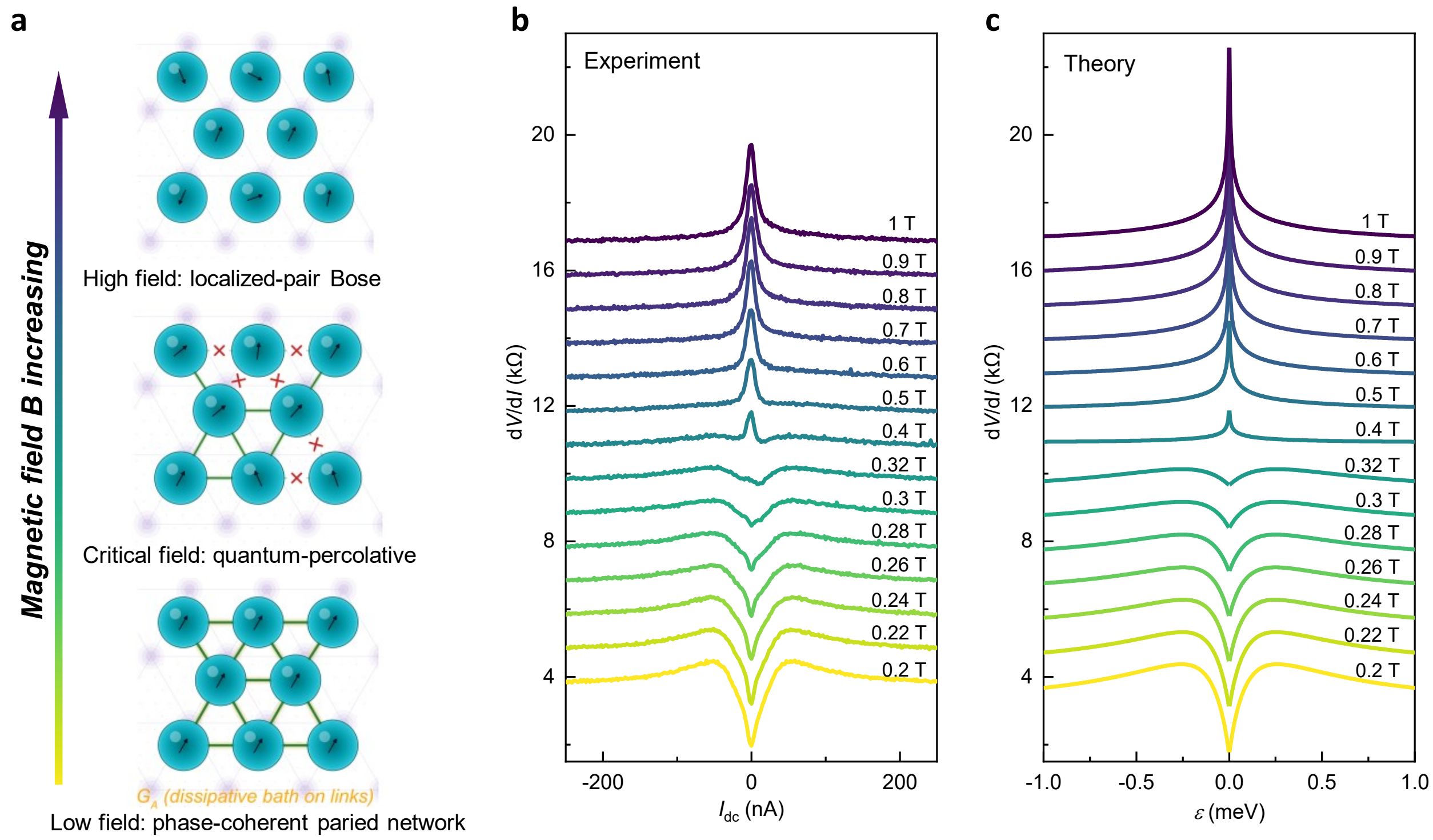

**Fig. 4 Microscopic processes of bosonic superconductor-insulator quantum phase transition. (a)** Schematic of the phase-network picture. At low field, local paired regions are connected by phase-coherent Josephson links protected by the Andreev subgap conductance $\mathcal{G}_A$. At the critical field, the coherent-link network loses global connectivity, with effective charging energy $E_C^* = E_C \exp[-\pi\mathcal{G}_A/4]$. At high field, local paired carriers survive but become phase-disordered and localized. Large dc bias or $T \gtrsim T^*$ drives the system into a common dynamic transport regime. **(b)** d$V$/d$I$ as a function of $I_{dc}$ for $B_\perp$ = 0.2-1 T at $T$ = 60 mK. The curves are vertically offset by 1 kΩ per trace for clarity. **(c)** Theoretical simulation of d$V$/d$I$ as a function of energy $\varepsilon$ for the magnetic field corresponding to those in **(b)**.